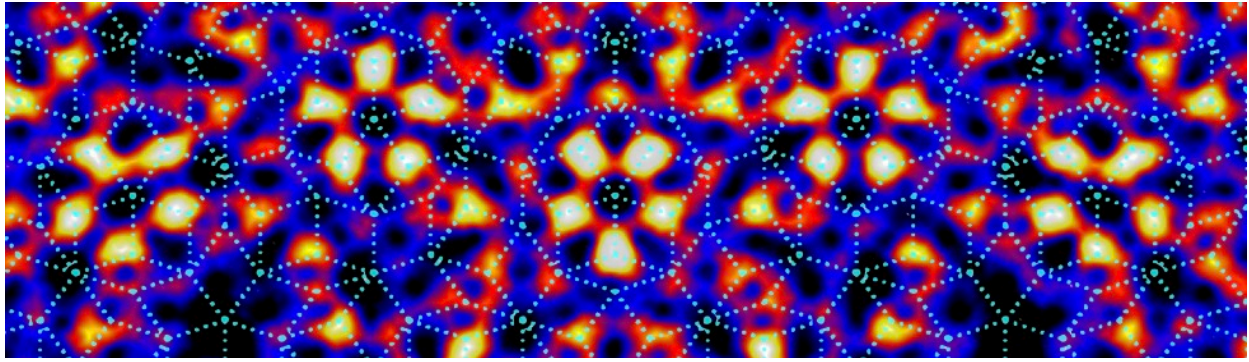

# Imaging Quasi-Periodic Electronic States in a Synthetic Penrose Tiling

Laura C. Collins, Thomas G. Witte, Rochelle Silverman, David B. Green, & Kenjiro K. Gomes[*]

*Department of Physics, University of Notre Dame, Notre Dame, Indiana 46556, USA*

*To whom correspondence should be addressed. E-mail: kgomes@nd.edu

**Abstract**

Quasicrystals possess long-range order but lack the translational symmetry of crystalline solids. In solid state physics, periodicity is one of the fundamental properties that prescribes the electronic band structure in crystals. In the absence of periodicity and the presence of quasicrystalline order, the ways that electronic states change remain a mystery. Scanning tunneling microscopy and atomic manipulation can be used to assemble a two-dimensional quasicrystalline structure mapped upon the Penrose tiling. Here, carbon monoxide molecules are arranged on the surface of Cu(111) one at a time to form the potential landscape that mimics the ionic potential of atoms in natural materials by constraining the electrons in the two-dimensional surface state of Cu(111). The real-space images reveal the presence of the quasiperiodic order in the electronic wave functions and the Fourier analysis of our results links the energy of the resonant states to the local vertex structure of the quasicrystal.

**Introduction**

The unique long range order of quasicrystals is evidenced by scattering[1]. Dan Shechtman first observed the ten-fold symmetric electron diffraction patterns from a rapidly cooled Al-Mn alloy[2]. Shechtman discovered a new type of material with long-range orientational and quasiperiodic order, but with rotational symmetries that can't coexist with the translational symmetry found in classically defined crystals[3]. The sharp scattering peaks are a result of this quasiperiodic order, which can be visualized by the formation of moiré patterns. This order arises from the repetition of the same tiling patterns over and over again, with well-defined rules albeit in a non-periodic manner[3]. The electronic behavior in periodic systems is dominated by extended wavefunctions while, in contrast, completely aperiodic or disordered systems are dominated by localized wavefunctions[4]. Quasicrystals tread right in the middle between these two conflicting limits, combining the lack of periodicity with the repetitions over long range, leading some studies to suggest critical wave functions which are neither localized nor extended[5].

The Penrose tiling is a classic example of quasiperiodic order and it has been studied by mathematicians even before the discovery of quasicrystals[6]. Tight-binding studies for Penrose vertex model quasicrystals with nearest neighbor edge hopping showed a largely degenerate state in the middle of the band containing almost 10% of all states in the band[7,8]. This degeneracy is typical for quasi-periodic tilings, and is due to families of strictly localized states[9] supported on a finite number of vertices which reappear at various places throughout the tiling[10]. These localized states may be caused not by a disordered local potential of the sites but instead by the local vertex structure that differentiates different sites[11]. The density of states of Penrose tiling quasicrystals is often described as spiky and presenting pseudogaps[12,13].

Photoemission measurements of natural quasicrystals have observed the existence of a pseudogap but they are not able to resolve the fine structure of the density of states[14,15]. Tunneling spectroscopic studies have also measured a pseudogap and they observe a spatially inhomogeneous local density of states, with signatures of electronic localization[16,17]. Even though high-quality imaging of quasicrystals has been achieved[18,19], many of the most pivotal theoretical predictions for the electronic properties in quasicrystals is yet to be to supported experimentally.

The use of synthetic quantum matter has been an extremely fruitful path to explore the physics behind natural systems. Synthetic quasicrystals have been modeled for cold atoms[20,21] and photonics[22]. In this experiment we realize a form of quantum simulation through the creation of artificial lattices[23,24]. In order to create a quasicrystal where we had unsurpassed control over the structure and imaging capabilities, we use the atomic manipulation capabilities of a scanning tunneling microscope (STM)[25] to assemble the quasicrystal one molecule at a time and spectroscopic imaging to visualize the electronic density of states.

**Results**

*Assembly and design of our synthetic quasicrystal*

Our experiment starts with a perfectly flat and clean Cu (111) surface, where we adsorbed carbon monoxide (CO) molecules. We used the STM tip to move each CO molecule into their correct positions prescribed by the rhombic Penrose tiling. This Penrose tiling is formed by the arrangement of two kinds of rhombi with the same side length but different vertex angles. One CO molecule is placed at the center of each rhombus of the tiling, repelling the electrons to the vertices of the tiles and allowing flow along their edges. This kind of arrangement of electrons at the vertices of the tiles and hopping along their edges is known as the Penrose vertex model quasicrystal and it has been extensively studied by tight-binding model calculations[26]. The STM topograph of the final assembled quasicrystal is shown in Fig. 1a with the overlay showing the Penrose tiling. The completed quasicrystal is made up of 460 CO molecules, spaced such that the length of the side of each rhombus measures $a_0 = 1.6$ nm. Since the CO molecules register to the position of Cu atoms on the surfaces, their placement is not exact for this non-periodic arrangement, but the imprecision is always smaller than 0.15 nm.

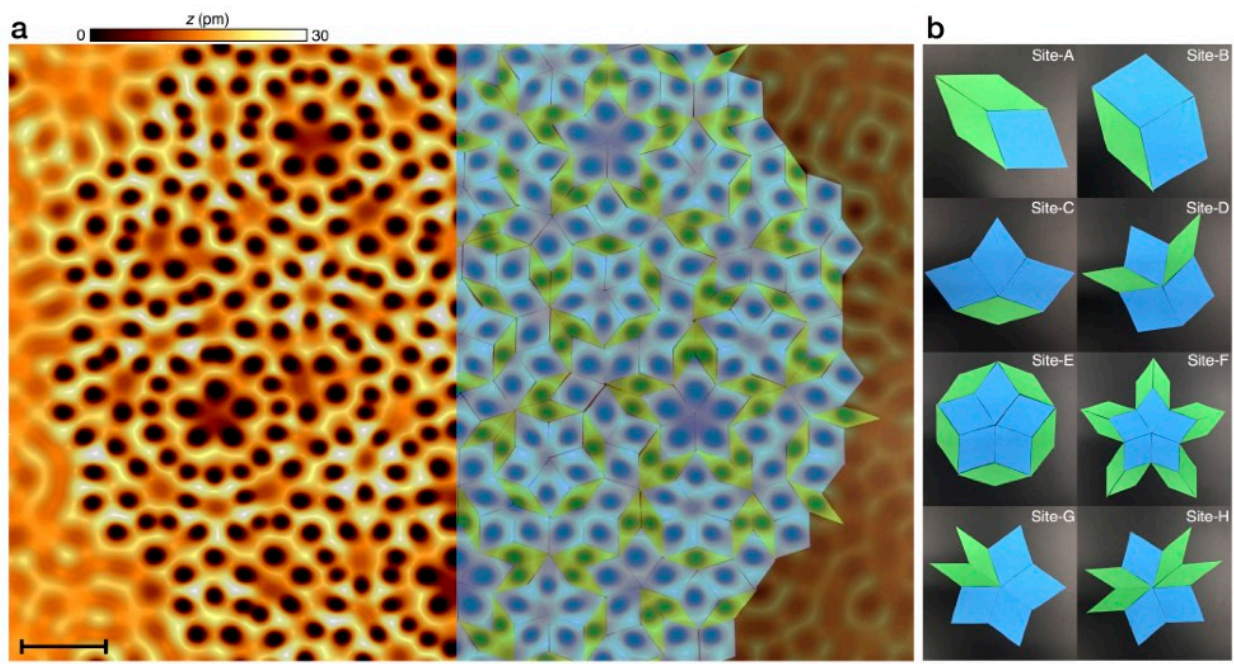

**Figure 1: Synthetic Penrose Tiling Quasicrystal** (a) STM topograph of assembled quasicrystal composed of 460 CO molecules measured at a bias voltage $V = 10$ mV and setpoint current $I = 1$ nA. The CO molecules are located at the center of each dark spot in the topograph. The overlay on the right side is the Penrose tiling composed of rhombi with side length $a_0 = 1.6$ nm and vertices angles 72°/108° (blue) and 36°/144° (green). Scale bar, 4 nm. (b) Atlas of the 8 types of vertex sites encountered in the Penrose vertex model tiling.

The Penrose tiling forms eight structurally different sites[27], associated with each type of first order vertex structure in the tiling, classified by the star combination of rhombi vertices. Interestingly, not every combination of rhombi is found in the Penrose tiling, as some configurations lead to spaces that cannot be extended. The eight types of sites allowed are shown in Fig. 1b, where the vertex at the center of each image is the vertex site. In our experiment, we aim to weight the role played by the vertex types prescribed by the quasicrystalline structure in determining the local density of states. Because each vertex site is formed by a different arrangement of CO molecules, we may end up with slightly different local potentials for each vertex site. However, two of the sites shown in Fig. 1b, site-E and site-F, are formed by the same union of 5 'fat' rhombi but are classified as structurally different[6]. Even though they are locally identical at first order, those sites are formed by different matching rules imposed by the global quasicrystalline order, and as a result they should exhibit different electronic behavior.

The complete quasicrystal is displayed in Fig. 2a. The visualization of the electronic states is done through the measurement of the differential tunneling conductance at different bias voltages, normalized by the spatially averaged tunneling conductance measured on the bare Cu surface (see Methods for details). The normalized conductance map taken at the Fermi energy of the Cu surface states (bias voltage $V = 0$ mV) reveals an intriguing pattern (Fig. 2b) where electrons form a standing wave resonance but only at particular sites. The histogram of the normalized conductance sorted by the site type reveals that the electrons are predominantly localized on the B-sites (Fig. 2c). The diagram in Fig. 2d highlights the position of the B-sites and makes the observed pattern even clearer upon a direct comparison with the conductance map. At this particular energy, electrons resonate at a particular local vertex structure of the quasicrystal sites but the existence of this unique state right at the Fermi energy is not accidental.

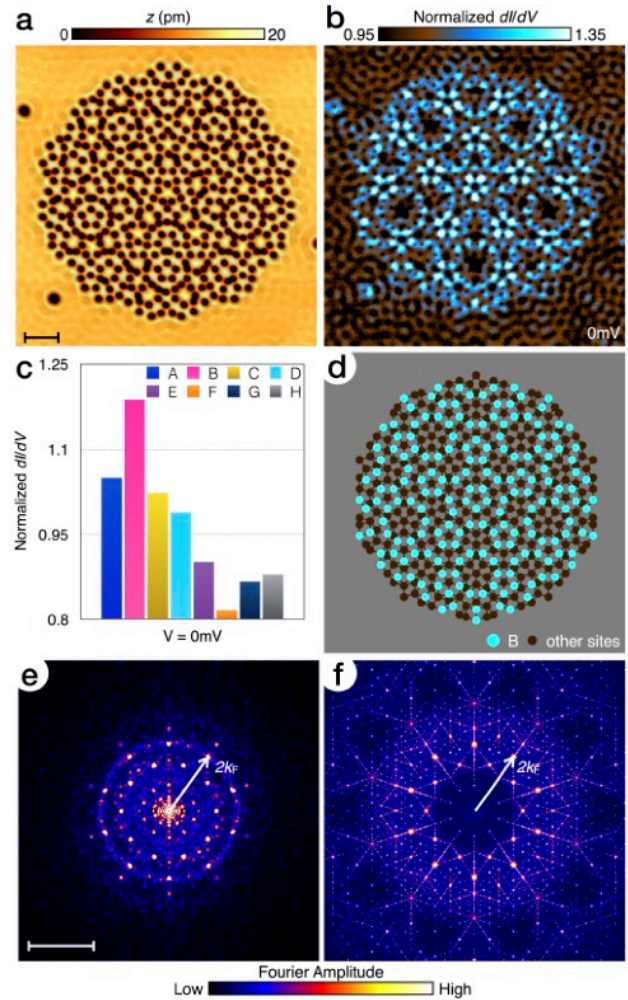

**Figure 2: Visualizing a Quasicrystalline Electronic State** (a) Topograph of the assembled quasicrystal composed of 460 CO molecules (dark spots in the image) measured with a bias voltage $V$ = 500 mV and setpoint current $I$ = 500 pA over a 42 nm by 42 nm field of view. Scale bar, 5 nm. (b) Normalized differential conductance map of the same field of view as (a) measured at bias voltage $V$ = 0 mV. (c) Histogram of the normalized differential conductance at bias voltage $V$ = 0 mV averaged over each type of vertex site. (d) Diagram of all vertex sites of the quasicrystal displaying the same field of view as (a) with site-B locations highlighted in cyan. (e) Fourier transform of the conductance map displayed in (b) with the arrow representing a wave-number value equal to twice the Fermi wave-number of the bare copper surface states ($k_F$). Scale bar, 4 nm$^{-1}$. (f) Fourier transform of quasicrystal model structure built with the same proportions as the one used in the experiment but with tens of thousands of sites to improve sharpness of Fourier peaks. The arrow represents the same wavenumber ($2k_F$) as in (e).

*Electronic behavior of our synthetic quasicrystal*

We have designed the tile size such that the position of the brightest peak in the Fourier transform of the Penrose tiling is equal to twice the Fermi wave-number ($k_F = 2.1$ nm$^{-1}$) of the electrons in the surface state of copper. In periodic lattices, this match corresponds to when the wave-vector first touches the edge of the Brillouin zone, transforming the band structure from electron-like to hole-like. This change in the curvature of the energy dispersion relation creates a nesting at the Fermi surface which leads to a resonance in the density of states which can be measured by tunneling spectroscopy. Fig. 2e is the Fourier transform of the conductance map shown in Fig. 2b. The outer ring originates from scattering of the electronic waves outside of the lattice and it has a radius of $2k_F$. Notice that it overlaps with the brightest peak of the Fourier transform of the model Penrose tiling (Fig. 2f). Quasicrystals don't have a well-defined Brillouin zone or dispersion relation but the Bragg peaks in the Fourier transform still correlate to the quasicrystalline order. The brightest peak is associated with the repetition of B-sites (the most abundant of all site types) even though this repetition is non-periodic, and seems to result in our observed standing wave resonance at those sites. We show additional normalized conductance maps and their Fourier transforms in Supplementary Note 1(Supplementary Fig. 1). To further clarify the connection between the quasi-periodic repetition of the B-sites and the brightest peak in the Fourier transform, we have taken the Fourier transform of a model Penrose tiling with only B-sites (Supplementary Fig. 2b) to examine the effects of their quasi-periodic arrangement. We see the same brightest peak as the Fourier transform with the full model Penrose tiling (Fig. 2f), which indicates that the chosen tile size for our Penrose tiling indeed leads to a resonant state due to the quasiperiodic nature of the tiling. A more detailed explanation can be found in Supplementary Note 2.

While the structural differences between these sites and the connection to the quasicrystalline order is clear, tight-binding model calculations on the Penrose tiling have found that sites that are structurally similar can present different local density of states[9]. The electronic behavior in our synthetic Penrose tiling is dictated by the Cu surface state electrons and their limited lifetime due to scattering to bulk states upon hitting the CO molecules. It is reasonable to assume that the electronic behavior of a site is due to its first order vertex structure. To prove this nearest-neighbor limit in our system, we performed further analysis of the B-sites separated according to their second order vertex structure[27] (Supplementary Fig. 3) and there was a very small variance between the behavior of the second order vertex structures. This analysis, along with a more detailed explanation can be found in Supplementary Note 2. Unlike tight-binding predictions[9] which predict the presence of forbidden B-sites at the Fermi energy, we see the same electronic behavior from each B-site in our synthetic quasicrystal system. We must emphasize though that the electronic states are not being determined simply by the local potential created by the CO molecules. We can demonstrate this by showing how the E and F-sites have distinct electronic behavior.

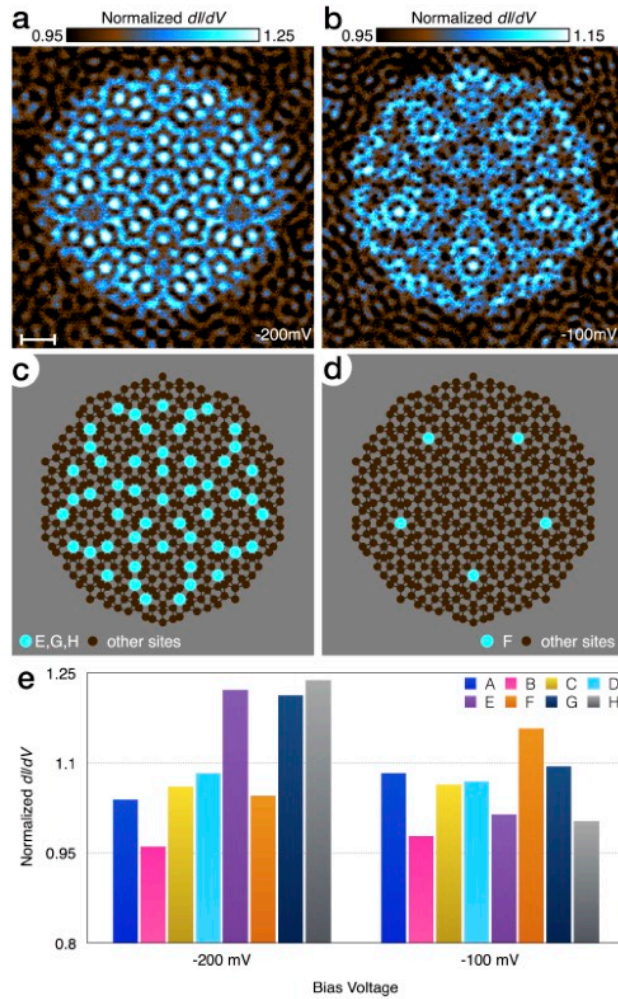

We measured conductance maps at different energies to further explore the dependence of resonant states in our synthetic Penrose tiling on the quasi-periodic arrangement of vertex sites. Conductance maps measured at bias voltage $V$ = -200 mV (Fig. 3a) and $V$ = -100 mV (Fig. 3b) show distinct patterns dominated by different site types. The map at $V$ = -200 mV is dominated by the E, G and H-sites (Fig. 3c) while the map at $V$ = -100 mV is dominated by the F-sites (Fig. 3d). The histograms of the conductance maps sorted by site type further demonstrate that pattern (Fig. 3e). Based on our comparison of the second order vertex structures of the B-sites (Supplementary Fig. 3) and the limited lifetime of the Cu surface state electrons, we assume the local potential of a site in our system is solely dependent on the arrangement of the nearest neighbors, so the E and F-sites have different local vertex structures but the same local potential since they are formed by the same pattern of five fat rhombi. We emphasize that even though the E and F-sites have the same local potential, the local density of states of these sites displays different energy dependence. The local density of states therefore depends on the arrangement of CO molecules beyond the first order vertex structure, which could reflect confinement effects from the second order vertex structure or the quasi-periodic arrangement of sites.

**Figure 3: Topologically Distinct Sites** (a) Normalized differential conductance map of a 42 nm by 42 nm field of view measured at bias voltage $V$ = -200 mV. Scale bar, 5nm. (b) Conductance map of the same field of view but at bias voltage $V$ = -100 mV. (c) Diagram of all vertex sites of the quasicrystal displaying the same field of view as above and with site-E, site-G and site-H highlighted in cyan. (d) Diagram of all vertex sites of the quasicrystal displaying the same field of view as above and with site-F highlighted in cyan. (e) Histogram of the normalized differential conductance at bias voltage $V$ = -200 mV (left) and $V$ = -100 mV (right) averaged over each type of vertex site.

Our next step to corroborate the dependence of the density of states on the local vertex structure was to measure the differential conductance spectra over a large sampling of sites. We measured 1/5 of the vertex sites up to a radius of 12 nm (Fig. 4a). We avoided the edges of the sample to minimize the finite size effects. We normalized each differential conductance spectrum in the same manner as we did with the maps, by dividing each spectrum by the spatially averaged spectrum of the bare copper surface (Fig. 4b). We compare this normalization method to a few others in

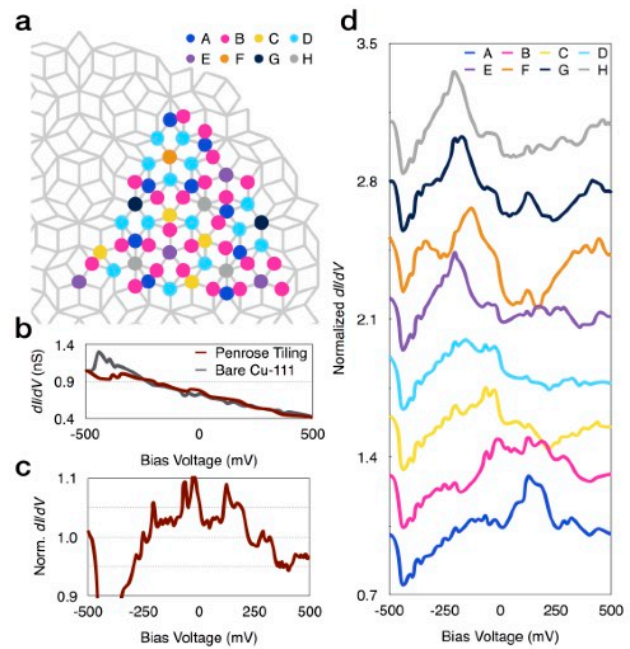

**Figure 4: Tunneling Spectroscopy Survey** (a) The gray diagram illustrates the Penrose tiling assembled in our experiment. The color dots mark the 61 electron sites where we measured the differential conductance spectra, with each color corresponding to a site type. (b) The red line is the average of all 61 differential conductance spectra measure in the quasicrystal and the gray line a differential conductance spectra of the bare Cu (111) surface, spatially averaged over 100 distinct points. (c) The total normalized conductance spectra, calculated by the ratio of the two spectra presented above. (d) Normalized conductance spectra averaged by site type for all eight sites in the Penrose tiling. The y-scale refers to the bottom spectra taken at the A-sites. Other spectra have been offset for clarity by 0.3 units in the y-scale above the previous spectrum.

Supplementary Note 3 (Supplementary Fig. 4). The total normalized conductance spectrum (Fig. 4c) can be obtained by averaging all spectra and it displays a spiky morphology, as expected for quasicrystals[28]. The abundance of peaks in the Fourier transform results in an abundance of gaps and resonances in the density of states. We note that our synthetic system has strong broadening effects in the conductance spectra because of the short lifetime of our surface state electrons due to scattering to the bulk states of Cu.

When comparing spectra taken at the same site types we noticed that they presented a very small variance between different locations. On the other hand, we noticed that spectra taken at different site types were consistently different, with a variance one hundred times larger than the variance of spectra taken at the same site types. In Fig. 4d, we present the spectra averaged by each site type. Notice that the sites with smaller areas (less distance between them and the nearest CO molecules in our system) show resonant states at higher energies than those with larger areas. These energy differences could be partially due to confinement effects in our synthetic Penrose tiling. However, since the E and F-sites have the same confinement at first order as each other and exhibit different resonant states (Fig. 3), both the local vertex structure and quasi-periodic order of the system are important in determining the electronic states.

**Discussion**

The creation of a synthetic Penrose tiling allowed for the first time to visualize the local density of state of a two-dimensional quasicrystal with sub-nanometer resolution. Our measurements have identified the existence of resonant states associated with Fourier peaks in reciprocal space and made a clear connection to the local vertex structure of each site. The extension of this experimental approach to include measurements of the changes in the electronic states due to scattering from defects could also solve other puzzles in the understanding of quasicrystals. Those measurements along with Fourier analysis should be able to reveal whether the electronic states are extended and dispersive or localized.

**Methods**

*Sample Preparation*

A single crystal Cu (111) sample was prepared by repeated cycles of sputtering (Ar ions, 1 keV) and annealing at a temperature of 1000 K in an ultra-high vacuum chamber. To deposit the CO molecules on the surface, we briefly exposed it to 1 x $10^{-8}$ Torr partial pressure of CO gas while keeping the sample temperature below 50 K.

*Normalized Differential Conductance Spectroscopy*

All differential conductance spectra and maps were measure with a lock-in amplifier technique, with set point voltage $V$ = -500 mV, set point current $I$ = 0.5 nA and bias voltage modulation with root-mean-square average of $dV$ = 3.5 mV. All energies are measured with the feedback loop open but the feedback loop is closed at the set point conditions to move between each point.

The differential conductance measured by a scanning tunneling microscope is proportional to the density of states of the assembled structure, but is also proportional to the density of states of the STM tip as well as the density of states of the bulk copper crystal. To isolate the density of states of the assembled lattice, we divided our measured conductance maps and tunneling spectroscopy by the spatially averaged differential conductance of the bare surface of copper. The bare copper differential conductance was measured with the same tip conditions as the other measurements, and it is averaged over 100 distinct sites. This normalization minimized the contribution of varying tip conditions and the copper background and ensured that we mainly considered electronic contributions from our assembled quasicrystal. We also find that the normalized spectra are a much better fit to theoretical models of tunneling spectroscopy, such as scattering theory.

**Data Availability**

The data that support the findings of this study are available from the corresponding author upon request.

**Author Contributions**

L.C.C. and K.K.G. conceived the project and designed the experiments. R.S. performed tight-binding calculations. T.G.W. and D.B.G. prepared the sample. T.G.W. and L.C.C. assembled the quasicrystal. L.C.C. and K.K.G. analyzed the data and wrote the paper.

**Competing Financial Interests**

The authors declare no competing financial interests.

## Supplementary Note 1

*Additional Conductance Maps and Fourier Transforms*

In Fig. 2 of the main text, we demonstrated that the electrons in our synthetic Penrose tiling quasicrystal form a standing wave resonance at the B-sites at the Cu surface states Fermi energy (bias voltage $V = 0$ mV). We have also shown that the brightest peak in the Fourier transform of the conductance map at this energy (map shown in Fig. 2b and again in Supplementary Fig. 1c; Fourier transform shown in Fig. 2e and again in Supplementary Fig. 1h) corresponds to the repetition of the B-sites. To further illustrate the connection between the bright peaks in the Fourier transform of the conductance map and the standing wave resonances at different energies and sites of our synthetic Penrose tiling quasicrystal, we include normalized differential conductance maps at several more energies along with their respective Fourier transforms in Supplementary Fig. 1. It is clear from these conductance maps that the electrons in the Cu surface states form standing wave resonances at different sites at different energies, and accordingly the bright peaks in the Fourier transforms of these maps highlight different resonances as well.

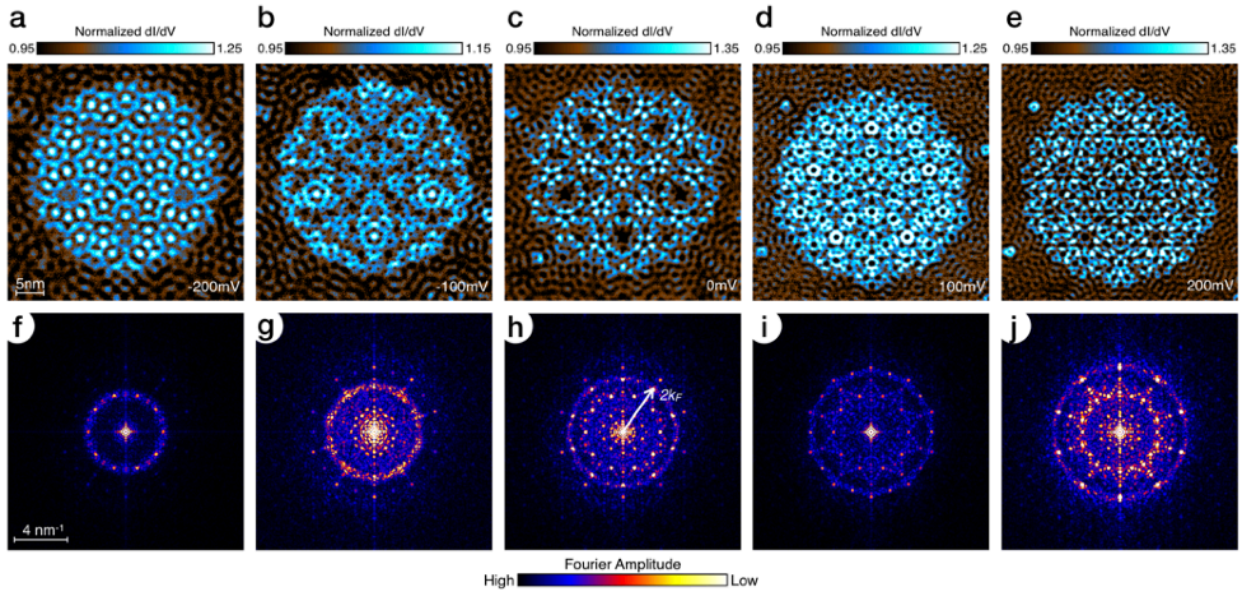

**Supplementary Figure 1: Visualizing More Quasicrystalline Electronic States** (a - e) Normalized differential conductance maps over a 42 nm by 42 nm field of view measured at bias voltages $V =$ -200 mV, -100 mV, 0 mV, 100 mV, and 200 mV respectively. Scale bar, 5nm. (f) Fourier transform of the conductance map displayed in (a). Scale bar, 4 nm$^{-1}$. (g) Fourier transform of the conductance map displayed in (b). (h) Fourier transform of the conductance map displayed in (c) with the arrow representing a wave-number value equal to twice the Fermi wave-number of the bare copper surface states ($k_F$). (i) Fourier transform of the conductance map displayed in (d). (j) Fourier transform of the conductance map displayed in (e).

**Supplementary Note 2**

*Extended Analysis of Electronic States Localized at the B-sites*

In Fig. 2f of the main text, we showed a Fourier transform of a model Penrose tiling with tens of thousands of sites (shown again in Supplementary Fig. 2a) to compare to the Fourier transform of our conductance map shown in Fig. 2e. To further explore the origin of the brightest peaks in the Fourier transform, we also took a Fourier transform of a model Penrose tiling with only the B-sites, since those are the sites that are the brightest in the conductance map at the Fermi energy (0 mV) and should therefore correspond to the brightest peaks in the Fourier transform. As you can see in Supplementary Fig. 2b, the brightest peaks in the Fourier transform of the model Penrose tiling with just the B-sites matches the brightest peaks in the Fourier transform of the model Penrose tiling with all of the sites. We also show the Fourier transform of a model Penrose tiling with just A-sites in Supplementary Fig. 2c, where we don't see bright peaks in the same positions as the other Fourier transforms, to demonstrate that this effect is not simply an artifact from excluding several of the sites in the Penrose tiling but comes from the quasi-periodic arrangement of the B-sites in the Penrose tiling. Therefore, this resonant state, evidenced by the bright peaks in the Fourier transform, corresponds to the quasi-periodicity of the B-sites.

In the main text, we claim that the different first order vertex structures can be distinguished electronically within our system. To prove this claim, we examine the electronic behavior of the B-sites further by analyzing them according to their second order vertex structure using the definitions shown in types for the nearest neighbor sites in Supplementary Fig. 3a. We have chosen to analyze the B-sites because in the sampling of sites we used to measure the local differential conductance spectra, shown in Fig. 4 of the main text, the B-sites were the most abundant. Also, in tight-binding calculations, it has been shown that B-sites exhibit different electronic behavior[2], and analyzing the B-sites in our system will allow us to directly compare the sites in our system with those in tight-binding systems. Once the B-sites were separated according to their second order vertex structures (Supplementary Fig. 3b), we normalized each spectrum by dividing by the spatially averaged spectrum of the bare copper surface, as described in the main text. In Supplementary Fig. 3c, we present the spectra averaged by each second order vertex structure. Notice that here we averaged together the spectra for the B6 site and the B7 site, since each second order vertex type only had one site and they are the mirror reflection of each other. In Fig. 4d, we showed the spectra for the different first order vertex structures and when we calculated the variance between each of those averaged spectra and the total density of states the variance was on the order of 0.1. When we calculated the variance between the averaged spectra for each second order vertex structure shown in Supplementary Fig. 3c and the average spectra of all the B-sites, the variance was on the order of $10^{-4}$. Since the variance in electronic behavior between the second order vertex structures is several orders of magnitude smaller than the variance between the first order vertex structures, we believe it is sufficient to say that the B-sites in our synthetic Penrose tiling system exhibit the same electronic behavior and we do not see forbidden B-sites which were predicted by tight-binding calculations[2].

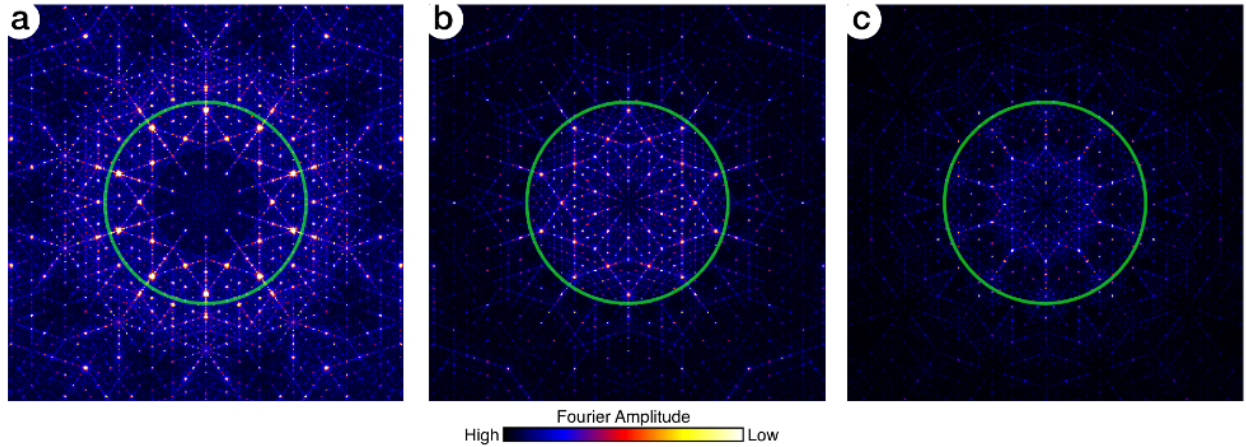

**Supplementary Figure 2: Fourier Transforms of Model Penrose Tilings** (a) Fourier transform of quasicrystal model structure built with the same proportions as the one used in the experiment but with tens of thousands of sites to improve sharpness of Fourier peaks. The overlaid cyan circle lies just outside a radius of $2k_F$ to highlight the brightest peaks, and is the same circle overlaid in (b) and (c).
(b) Fourier transform of a model Penrose tiling with only the B-sites, again using tens of thousands of sites to improve sharpness, with a cyan circle overlaid just outside a radius of $2k_F$ to highlight the brightest peaks. (c) Fourier transform of a model Penrose tiling with only A-sites, using tens of thousands of sites with a cyan circle overlaid just outside a radius of $2k_F$ to compare to the other Fourier transforms.

Supplementary Fig. 3a[1]. To make these definitions easier to visualize, we have also provided the site

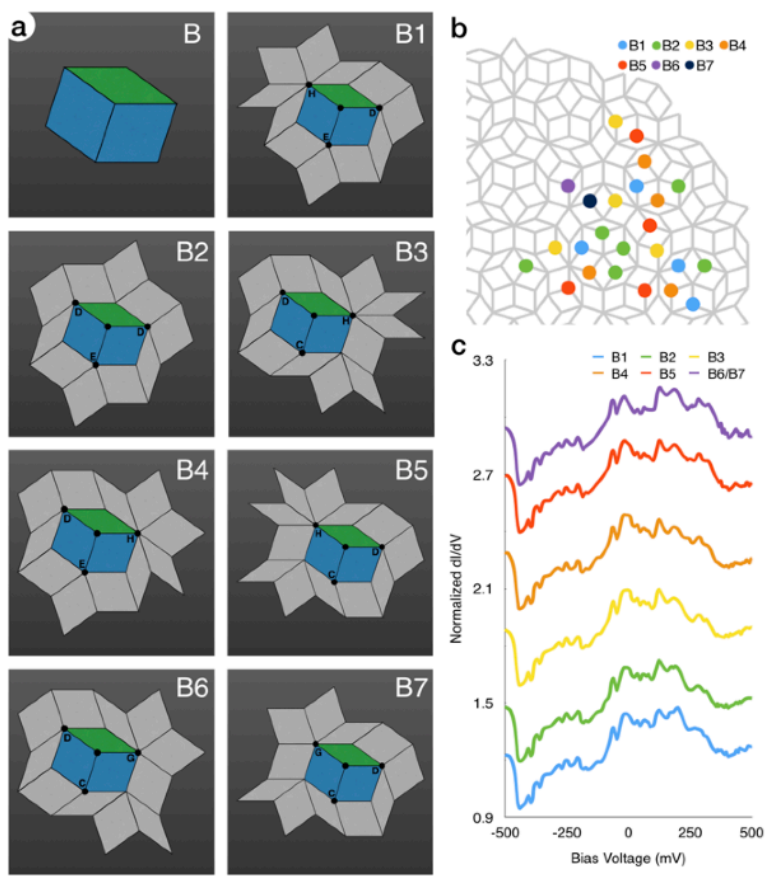

**Supplementary Figure 3: Second Order Vertex Structure for the B-sites** (a) The 7 possible second order vertex structures for B-sites with nearest neighbor site types labeled. (b) The gray diagram illustrates the Penrose tiling assembled in our experiment. The color dots mark the classification of each B-site where we measured the differential conductance spectra. (c) Normalized conductance spectra averaged by second order vertex type (where B6 and B7 have been averaged together since they are mirror images of each other, and there is only one spectrum for each type). The y-scale refers to the bottom spectra taken at the B1-sites. Other spectra have been offset for clarity by 0.4 units in the y-scale above the previous spectrum.

**Supplementary Note 3**
*Normalization of Differential Conductance*

As discussed in the main text, we measured the differential conductance spectra of our synthetic Penrose tiling over a large sampling of sites. The spectra shown in the main text were all normalized by dividing each spectrum by the spatially averaged spectrum of the bare copper surface, shown in Fig. 4b. To demonstrate that the features we see in these spectra are not artifacts of the normalization method we use, we present selected spectra using another normalization method in Supplementary Fig. 4. Supplementary Fig. 4a is the same result from Fig. 4c in the main text shown between -300 mV and 300 mV to make it easier to see the individual peaks and features, and is included here for simpler comparison between the two normalization methods. The spectra in Supplementary Fig. 4b-c were normalized by

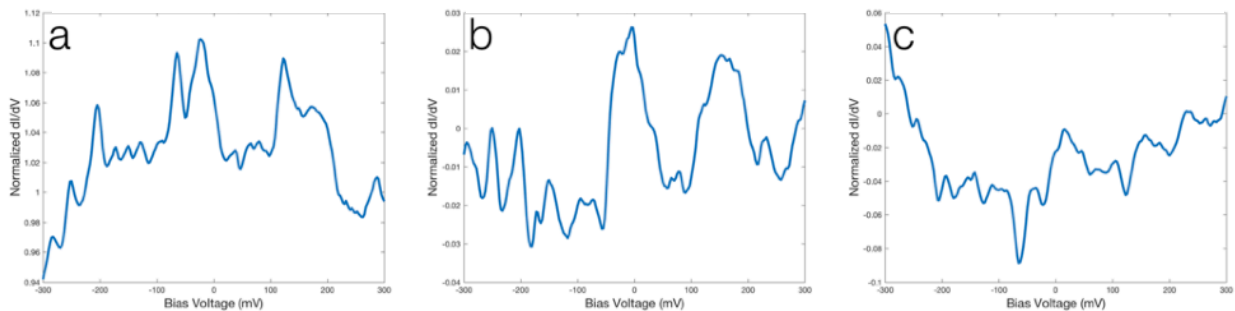

**Supplementary Figure 4: Normalizing the Tunneling Spectroscopy a Different Way** (a) The total normalized conductance spectra, calculated by the ratio of the average of all 61 differential conductance spectra measured in the quasicrystal and the spatially averaged differential conductance spectra of the bare Cu surface, as shown in Fig. 4c. (b) The total normalized conductance spectra, calculated by subtracting a linear fit of the spatially averaged bare Cu differential conductance spectra from the average of all 61 differential conductance spectra measured in the quasicrystal. (c) The total normalized conductance spectra of the bare Cu, calculated by subtracting a linear fit of the bare Cu differential conductance spectra from the spatially averaged bare Cu differential conductance spectra.

subtracting a linear fit of the spatially averaged spectrum of the bare copper surface from the total normalized conductance spectrum and the spatially averaged spectrum of the bare copper surface, respectively. We can see in Supplementary Fig. 4c that the bare copper spectrum is not exactly linear, since there are still some slight peaks in the resultant spectrum. More importantly, when we compare

Supplementary Fig. 4a and 4b, we can see that while some of the smaller peaks and features have different magnitudes after each normalization and some of the larger peaks in Supplementary Fig. 4a are harder to distinguish from each other in Supplementary Fig. 4b. However, since both methods show significant peaks in the same locations, we conclude that the features we see are not due to the normalization method we have used in the main text.

**Supplementary References**